\documentclass[preprint,aps,prb,11pt]{revtex4-1}

\usepackage{amsmath}
\usepackage{braket}
\usepackage{graphicx}
\usepackage{multirow}

\begin{document}

\title{Numerical determination of a non-equilibrium many-body statistical operator for quasi-bound electrons in a gated nanowire system}

\author{J. M. Castelo and K. M. Indlekofer}

\affiliation{Hochschule RheinMain University of Applied Sciences\\ IMtech / Faculty of Engineering\\ D-65428 R\"usselsheim, Germany}

\date{17.06.2014}

\begin{abstract}
We present a numerical approach to construct a non-equilibrium many-body statistical operator $\hat{\rho}_\mathrm{rel}$ for an adaptive subspace of relevant quasi-bound electronic states in a semiconductor nanowire-based field-effect transistor (NWFET). As a constraint for $\hat{\rho}_\mathrm{rel}$, we assume that the single-particle density matrix $\rho_1$ is a given quantity, resulting from a non-equilibrium Green's function (NEGF) calculation for the NWFET for a given set of applied voltages. Two different orthonormal (ON) eigenbases for $\hat{\rho}_\mathrm{rel}$ are considered: (A) a Slater determinant basis of natural orbitals (eigenstates of $\rho_1$) and (B) the eigenbasis of the projected many-body Hamiltonian $\hat{H}_\mathrm{rel}$ within a relevant Fock subspace of the system. As for the eigenvalues $w_n$ of $\hat{\rho}_\mathrm{rel}$, we furthermore assume that $w_n$ have a generalized Boltzmann form, parameterized by effective electrochemical potentials of natural orbitals and a given temperature. From the determined $\hat{\rho}_\mathrm{rel}$, in turn, one can calculate expectation values for any many-body observable within the relevant subspace. As an example, we analyze the electron density and the covariance of the density-density correlation function for representative electronic preparations of the NWFET.
\end{abstract}

\maketitle

\section{Introduction}
\label{sec:introduction}

As the downscaling process of the channel length of field-effect transistors (FETs) reaches the submicron limit, several difficulties emerge in the construction of a functionally well-behaved device~\cite{chaudhry2013}. One of these matters is the necessity of a proper gate control over the carriers. The electrostatic integrity of the device must be maintained, the gate potential rather than the drain potential must control the charge state inside the channel. Otherwise, so-called short-channel-effects may appear. It can be seen~\cite{appenzeller2008} that these effects may be avoided by employing a one-dimensional (1D) semiconductor nanowire as the FET channel, provided that the reduction in channel length is accompanied by a corresponding scaling of nanowire diameter and oxide thickness.  In this respect, nanowires are ideally suited for ultimately scaled FET devices because of their cylindrical shape with a scalable diameter into the few nanometer range. They constitute the basis of the nanowire field-effect transistor (NWFET)~\cite{dey2012,larrieu2013}.

A successful quantum kinetic simulation method to study non-equilibrium electronic transport in nanoscale devices is the non-equilibrium Green's function (NEGF) approach~\cite{datta1999,rammer2007,haug2008}. Usually, NEGF methods employ a mean-field approximation to model electron--electron Coulomb interaction. But for ever-decreasing channel lengths, the number of electrons involved in the on-state current is small and few-electron Coulomb blockade~\cite{kastner1992,likharev1999,bjork2004} effects start to play a role, which cannot be taken into account by a mean-field approximation. Thus, to describe these, a many-body formulation of Coulomb interaction is needed. The dimension of the full Fock space grows exponentially with the number of single-particle states, therefore a full many-body treatment of this interaction is computationally unfeasible for single-particle bases with $\gtrsim10$ states, as required by realistic systems. Nevertheless, it is possible to reduce the degrees of freedom to a set of relevant ones, self-consistently and adaptively, as the multi-configurational self-consistent Green's function (MCSCG) method proposes~\cite{indlekofer2005,indlekofer2007}, thus rendering a many-body description of electron--electron Coulomb interaction numerically feasible.

Various authors~\cite{hershfield1993,bokes2003,dhar2012,ness2013} have considered the general mathematical construction of a non-equilibrium many-body statistical operator of interacting electrons for given external constraints or bias conditions. In this paper, we present an adaptive numerical approach to determine a reduced non-equilibrium many-body statistical operator $\hat{\rho}_\mathrm{rel}$ for quasi-isolated electronic states within the channel of a realistic NWFET system. The underlying physical model assumes the knowledge of the (self-consistent) single-particle density matrix~\cite{lowdin1955} $\rho_1$ of the whole channel system for the given gate and bias voltage condition. In turn, the single-particle Hilbert space of the whole channel system is divided into a small, adaptively chosen relevant subspace and an orthogonal rest, following the idea of the MCSCG approach. Here, relevant basis states are defined as natural orbitals (i.e., eigenstates of $\rho_1$)~\cite{lowdin1955,davidson1972} which are quasi-isolated (i.e., resonantly trapped) and exhibit occupation fluctuations (i.e., being neither empty nor fully occupied), thus being responsible for few electron Coulomb blockade~\cite{kastner1992,likharev1999,bjork2004} effects. The latter subspace requires a Fock space treatment of the Coulomb interaction, beyond the commonly employed mean-field approximation, whereas interaction terms of the orthogonal rest are treated by a conventional mean-field approximation. In the present case, $\rho_1(V_\mathrm{GS},V_\mathrm{DS})$ is determined self-consistently from a NEGF calculation of non-equilibrium electronic transport in the NWFET channel for given gate and bias voltages $V_\mathrm{GS}$ and $V_\mathrm{DS}$, respectively. In turn, a $\rho_1$-adaptive relevant Fock subspace is defined, constructed from relevant natural orbitals as defined above. From the given matrix $\rho_1$, in turn, a reduced many-body statistical operator $\hat{\rho}_\mathrm{rel}$ within the relevant subspace can be constructed. Here, the given matrix elements of $\rho_1$ impose constraints on $\hat{\rho}_\mathrm{rel}$. In comparison, the approach described in Ref.~\cite{bokes2003} is based on direct constraints on single-particle observables (such as the electronic current), whereas the approach described in this paper is based on a general $\rho_1$ with an adaptive relevant Fock subspace. Finally, with the help of $\hat{\rho}_\mathrm{rel}$, expectation values of any observable (and correlation functions) of relevant states can be calculated numerically.

In general, the Fock subspace operator $\hat{\rho}_\mathrm{rel}$ is not uniquely defined by the constraint of a given single-particle matrix $\rho_1$. Further physical assumptions are therefore required. In the present paper, we assume that the eigenvalues $w_n$ of $\hat{\rho}_\mathrm{rel}$ are of a generalized grand-canonical Boltzmann form (to maximize entropy), parameterized by a set of effective electrochemical potentials and an effective temperature. Furthermore, for the assumed many-body eigenbasis of $\hat{\rho}_\mathrm{rel}$, two alternatives are considered in this paper: (A) Slater determinants of relevant natural orbitals and (B) the eigenbasis of the projected many-body Hamiltonian within the relevant subspace. In order to determine an optimum $\hat{\rho}_\mathrm{rel}$ that satisfies the given constraints on $\rho_1$ numerically, a genetic algorithm is employed that searches for the optimum solution that minimizes a suitably defined deviation measure.

The organization of the paper is as follows: In Sec.~\ref{sec:statistical_operator}, we explain in detail how the numerical determination of the statistical operator is performed. In Sec.~\ref{sec:examples}, examples are presented of expectation values of observable quantities that can be obtained from the statistical operator, such as the electron density and the density-density correlation function (and the resulting covariance). Finally, we give a conclusion in Sec.~\ref{sec:conclusions}.

\section{Numerical determination of the statistical operator}
\label{sec:statistical_operator}

\subsection{Single-particle density matrix}

The single-particle density matrix~\cite{lowdin1955} $\rho_1$ of the system can be obtained by means of the NEFG formalism. Within the employed NEGF approach, the nanowire channel is described as a 1D single-band tight-binding chain in the effective mass approximation, represented by a localized orbital ON basis with $N_{\mathrm{max}}=2\times N_{\mathrm{sites}}$ spin/site orbitals, where the factor 2 stems from spin degree of freedom and $N_{\mathrm{sites}}$ denotes the number of localized spatial sites. In this representation, $\rho_1$ is obtained as follows
\begin{equation}
 \rho_{1_{jk}} = \frac{1}{2\pi i} \int dE ~ G^{<}_{jk}(E)
\end{equation}
where $G^{<}$ is the energy dependent lesser Green's function~\cite{datta1999,rammer2007,haug2008} in matrix form and the indices correspond to localized spin/site orbitals. Here, $G^{<}$ and the relevant subspace is determined self-consistently by means of the NEGF/MCSCG formalism~\cite{indlekofer2005,indlekofer2007}.

The dimension of the matrix $\rho_1$ reads as $N_{\mathrm{max}}\times N_{\mathrm{max}}$. Its eigenvectors are known as natural orbitals and its eigenvalues $\xi_i$ can be interpreted as average occupation numbers of these states~\cite{lowdin1955,davidson1972}. They satisfy $0 \leq \xi_i \leq 1$. If $U$ denotes the unitary transformation matrix that diagonalizes $\rho_1$, such that $\rho_1^{\mathrm{diag}}=U^\dagger\rho_1 U$ is the single-particle density matrix in diagonal form,  the natural orbitals are represented by the columns of $U$.

\subsection{Relevant Fock subspace}

Given that the single-particle ON basis has $N_\mathrm{max}$ states, the resulting many-body Fock space $\mathcal{F}$ has dimension $\mathrm{dim}(\mathcal{F})=2^{N_\mathrm{max}}$. The set of all Slater determinants of ON natural orbitals constitute an ON basis of the whole Fock space, corresponding to states with well defined occupation (0 or 1) of single-particle basis states for the chosen basis of natural orbitals. Thus, a Slater determinant $\Ket{D}$ can be uniquely identified in the occupation number representation by a vector of $N_\mathrm{max}$ bits $b_i\in\{0,1\}$ of the form $\Ket{D}=\Ket{b_1b_2\cdots b_{N_\mathrm{max}}}$.

\begin{figure}[tb]
 \centering
 \includegraphics[width=0.85\linewidth]{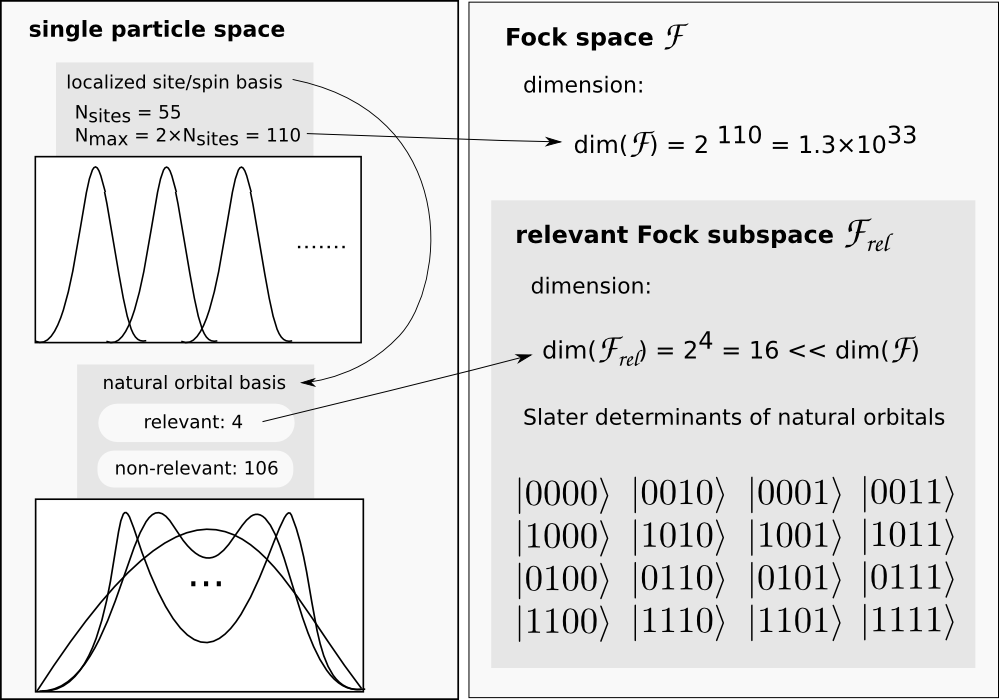}
 \caption{Example of the construction of the relevant many-body Fock subspace after the identification of relevant single-particle states. Sketches of the single-particle localized spin/site and natural orbitals bases are shown.}
 \label{fig:relsubspace}
\end{figure}

To make calculations of realistic nanowire devices numerically feasible, instead of considering the full Fock space, we restrict ourselves to a relevant subspace $\mathcal{F}_\mathrm{rel}$. The Slater determinants $\Ket{D}$ of $\mathcal{F}_\mathrm{rel}$ are constructed as follows (see also Fig.~\ref{fig:relsubspace}). There are as many Slater determinants as possible bit combinations of zeros and ones (empty and occupied states) of the $N_\mathrm{rel}$ relevant natural orbitals. Here, ``relevant'' is defined as being fluctuating (i.e. $0<\xi_i<1$ with given thresholds) and weakly coupled to the contacts (i.e. the magnitude of the imaginary part of the contact coupling selfenergy is below a given threshold). Thus, the dimension of the relevant Fock subspace is $\mathrm{dim}(\mathcal{F}_\mathrm{rel})=2^{N_\mathrm{rel}}$. The natural orbitals whose average occupation numbers are close to unity ($\xi_i\simeq1$) can be treated in two alternative ways: (i) they may be treated as being fully occupied (bit set to 1 for $N_\mathrm{occ}$ bits) in every $\Ket{D}$ and thus being incorporated statically within the many-body Fock subspace, or (ii) they may be set as empty (bit set to 0) in every $\Ket{D}$ and being incorporated in a mean-field way (in the single-particle part of the Hamiltonian). The rest of the natural orbitals (i.e., those which are empty with $\xi_i\simeq0$ or which are fluctuating but being strongly coupled to the contacts) are kept empty (bit set to 0 for  $N_\mathrm{rest}$ bits) in every $\Ket{D}$ and being incorporated in a mean-field way (in the single-particle part of the Hamiltonian). The same reasoning holds for the calculation of expectation values of general observables, containing contributions from the relevant subspace (where $\rho_1$ is known and $\hat{\rho}_\mathrm{rel}$ will be determined) and the rest (where only $\rho_1$ is known).

As an example (see Fig.~\ref{fig:relsubspace}), for $N_\mathrm{max}=110$, the dimension of the Fock space is $\mathrm{dim}(\mathcal{F})\simeq1.3\times10^{33}$, making a full many-body approach to electron--electron interaction unfeasible for typical lengths of the NWFET. On the other hand, if only $N_\mathrm{rel}=4$ natural orbitals are relevant then the dimension is $\mathrm{dim}(\mathcal{F}_\mathrm{rel})=16$ and the many-body approach becomes numerically feasible.

\subsection{Projected many-body Hamiltonian}

The many-body Hamiltonian $\hat{H}$ of the total system is composed of single-particle $\hat{H}_0$ and electron--electron Coulomb interaction $\hat{H}_{ee}$ terms. It has the form~\cite{indlekofer2005}
\begin{equation}
 \hat{H}=\hat{H}_0+\hat{H}_{ee} = \sum\limits_{i,j=0}^{N_\mathrm{max}-1}h_{ij}\tilde{c}^\dagger_i \tilde{c}_j + \frac{1}{2}\sum\limits_{i,j,k,l=0}^{N_\mathrm{max}-1}V_{ijkl}\tilde{c}^\dagger_i\tilde{c}^\dagger_j\tilde{c}_k\tilde{c}_l
\end{equation}
where $\tilde{c}^\dagger_i$ and $\tilde{c}_i$ (with tilde) are the creation and annihilation operators for the localized spin/site orbital basis states (where the spin is included implicitly in the single-particle indices). 

The projected many-body Hamiltonian $\hat{H}_\mathrm{rel}$ within the relevant Fock subspace $\mathcal{F}_\mathrm{rel}$ is obtained from $\hat{H}$ via projection to the relevant Fock subspace $\mathcal{F}_\mathrm{rel}$. In order to obtain the matrix elements $H^\mathrm{D}_{\mathrm{rel}_{nm}}$ of $\hat{H}_\mathrm{rel}$ with respect to the ON basis of Slater determinants $\Ket{D_m}$ of natural orbitals, two kinds of operator products need to be evaluated: single-particle terms $\Bra{D_n}c^\dagger_i c_j\Ket{D_m}$ and two-particle terms $\Bra{D_n}c^\dagger_i c^\dagger_j c_k c_l\Ket{D_m}$. Here, $c_i^\dagger$ and $c_i$ (without tilde) represent creation and annihilation operators for the natural orbital basis states.

Once the matrix representation of $\hat{H}_\mathrm{rel}$ is obtained, it can be diagonalized, yielding $\mathrm{dim}(\mathcal{F}_\mathrm{rel})$ eigenstates $\Ket{\psi_n}$ and eigenenergies $\epsilon_n$ that satisfy the eigenvalue equation $\hat{H}_\mathrm{rel}\Ket{\psi_n}=\epsilon_n\Ket{\psi_n}$. Every energy eigenvector can be expressed in the ON basis of Slater determinants as
\begin{equation}
 \Ket{\psi_n} = \sum\limits_{i=0}^{\mathrm{dim}(\mathcal{F}_\mathrm{rel})-1} \Lambda_{in} \Ket{D_i}
\end{equation}
where $\Lambda_{in}=\Braket{D_i|\psi_n}\in\mathcal{C}$, which defines a unitary transformation matrix $\Lambda$ that diagonalizes the projected many-body Hamiltonian matrix: $\Lambda^\dagger H^\mathrm{D}_\mathrm{rel}\Lambda=H^\mathrm{diag}_\mathrm{rel}$.

\subsection{Many-body statistical operator}

The statistical preparation of the relevant subsystem is given by the reduced many-body statistical operator 
\begin{equation}
\label{eq:rho_mb}
 \hat{\rho}_\mathrm{rel} = \sum_n w_n \Ket{\phi_n}\Bra{\phi_n} \hspace{5mm}. 
\end{equation}
It is Hermitian and satisfies $\mathrm{Tr}(\hat{\rho}_\mathrm{rel})=\sum\limits_n w_n = 1$, with $w_n\in[0,1]$. The eigenvalue $w_n$ can be interpreted as the probability associated to the eigenstate $\Ket{\phi_n}$. In order to determine $\hat{\rho}_\mathrm{rel}$, we require that its probability distribution maximizes the entropy subject to certain constraints. This determines the functional form of the eigenvalues $w_n$ of $\hat{\rho}_\mathrm{rel}$.

So far, no assumption has been made about the eigenvectors $\Ket{\phi_n}$ that form the ON eigenbasis of $\hat{\rho}_\mathrm{rel}$ within the relevant Fock subspace $\mathcal{F}_\mathrm{rel}$. In the following, we consider two cases: (A) the Slater determinant basis of natural orbitals and (B) the eigenbasis of $\hat{H}_\mathrm{rel}$ within the relevant subspace. One has to note that for a stationary relevant subsystem, where the time evolution of $\hat{\rho}_\mathrm{rel}$ is assumed to be driven solely by the projected $\hat{H}_\mathrm{rel}$, both operators need to commute, leading to the basis choice (case B) where $\Ket{\phi_n}$ are chosen as eigenstates of $\hat{H}_\mathrm{rel}$.

\subsection{Single-particle density matrix constraint}

The constraint that $\hat{\rho}_\mathrm{rel}$ must satisfy is given by the following expression
\begin{equation}
\label{eq:rho1_constraint}
 \rho_{1_{ij}} \overset{!}{=} \mathrm{Tr}(\hat{\rho}_\mathrm{rel}c^\dagger_j c_i)
\quad
(\forall i,j \in\,\mathrm{relevant})
\end{equation}
which links $\hat{\rho}_\mathrm{rel}$ with the given $\rho_1$ within the $N_\mathrm{rel}$ dimensional subspace of relevant natural orbitals. It results from the fact that the expectation value of any single-particle observable $\hat{A}=\sum\limits_{i,j=0}^{N_\mathrm{max}-1}a_{ij}c^\dagger_ic_j$ can be obtained in the following two equivalent ways
\begin{equation}
 \Braket{\hat{A}} = \mathrm{Tr}(\hat{\rho}_\mathrm{rel}\hat{A}) = \sum_{i,j=0}^{N_\mathrm{max}-1}a_{ij}\mathrm{Tr}(\hat{\rho}_\mathrm{rel}c^\dagger_ic_j)
\end{equation}
\begin{equation}
 \Braket{\hat{A}} = \mathrm{Tr}(\rho_1a) = \sum_{i,j=0}^{N_\mathrm{max}-1}a_{ij}\rho_{1ji} \hspace{5mm}.
\end{equation}
Within the relevant single-particle subspace ($N_\mathrm{rel}$ dimensional), equation~\eqref{eq:rho1_constraint} provides a set of $N_\mathrm{rel}\times N_\mathrm{rel}$ complex conditions. Noting that both sides of the equation are Hermitian, we can see that the number of unique real constraints is reduced to $N_\mathrm{rel}\times N_\mathrm{rel}$.

Since we are using the natural orbital basis in which the given $\rho_1$ is diagonal with eigenvalues $\xi_i$ and according to the expression of $\hat{\rho}_\mathrm{rel}$ determined by Eq.~\eqref{eq:rho_mb}, the constraints imposed by Eq.~\eqref{eq:rho1_constraint} can be rewritten as
\begin{equation}
\label{eq:rho1_constraint_expanded}
 \xi_i\delta_{ij} \overset{!}{=} \sum_n w_n \Bra{\phi_n}c^\dagger_j c_i\Ket{\phi_n}
\quad
(\forall i,j \in\,\mathrm{relevant}).
\end{equation}
Note that $i$ and $j$ are restricted to the indices of the relevant natural orbital basis states.

One has to note that the diagonal constraints ($i=j$) are equivalent to constrained average particle numbers $\Braket{\hat{n}_i}$ for the corresponding natural orbitals. For the following subsections, we denote the resulting coefficients as
\begin{equation}
\label{eq:diagonal_coeffs}
 \nu_i^n \equiv \Bra{\phi_n}c^\dagger_i c_i\Ket{\phi_n} = \Bra{\phi_n}\hat{n}_i\Ket{\phi_n}
\end{equation}
so that this subset of $\rho_1$-diagonal constraints read as
\begin{equation}
\label{eq:rho1_diagonal_constraint}
 \xi_i \overset{!}{=} \sum_n w_n \nu_i^n
\quad
(\forall i \in\,\mathrm{relevant}).
\end{equation}

\subsection{Case A: Slater determinant basis of natural orbitals}

In this case, the statistical operator corresponds to a mixture of Slater determinants of relevant natural orbitals, that is $\lbrace \Ket{\phi_n} \rbrace=\lbrace \Ket{D_n} \rbrace$. Thus, we obtain
\begin{equation}
 \hat{\rho}_\mathrm{rel} = \sum_{n=0}^{\mathrm{dim}(\mathcal{F}_\mathrm{rel})-1} w_n \Ket{D_n}\Bra{D_n} \hspace{5mm}.
\end{equation}
In the Slater determinant basis of natural orbitals $\lbrace \Ket{D_n} \rbrace$, its matrix representation is diagonal. The off-diagonal constraints given by Eqs.~\eqref{eq:rho1_constraint_expanded} vanish. The remaining equations Eqs.~\eqref{eq:rho1_diagonal_constraint} for the diagonal terms read as 
\begin{equation}
\label{eq:constraints_A}
 \xi_i \overset{!}{=} \sum_{n=0}^{\mathrm{dim}(\mathcal{F}_\mathrm{rel})-1} w_n N_i(\Ket{D_n})
\quad
(\forall i \in\,\mathrm{relevant}),
\end{equation}
where we defined $N_i(\Ket{D_n}) \equiv \Bra{D_n}\hat{n}_i\Ket{D_n} = \nu_i^n$ as the occupation number of the relevant natural orbital $i$ for the Slater determinant $\Ket{D_n}$. The energies associated with each eigenvector $\Ket{D_n}$ of $\hat{\rho}_\mathrm{rel}$ are given by the diagonal elements of $H^\mathrm{D}_\mathrm{rel}$ and read as $E_n = \Bra{D_n}\hat{H}_\mathrm{rel}\Ket{D_n}=H^\mathrm{D}_{\mathrm{rel}_{nn}}$.  Both quantities, $\nu_i^n$ and $E_n$ are required in Eqs.~\eqref{eq:boltzmann} and \eqref{eq:partition} below.

One has to note that the vector $\mathbf{w}$ of eigenvalues of $\hat{\rho}_\mathrm{rel}$ with $\mathrm{dim}(\mathcal{F}_\mathrm{rel})=2^{N_\mathrm{rel}}$ components has to satisfy a set of $N_\mathrm{rel}$ conditions given by Eqs.~\eqref{eq:constraints_A}, so the number of unknowns is greater than the number of constraints and the problem is underdetermined in general, thus requiring further assumptions about the function form of $w_n$, such as a Boltzmann form (see below).

\subsection{Case B: eigenbasis of $\hat{H}_\mathrm{rel}$}

In this case, we assume a stationary state of a relevant subsystem that is solely driven by the projected $\hat{H}_\mathrm{rel}$. Thus, we have $[\hat{\rho}_\mathrm{rel},\hat{H}_\mathrm{rel}]=0$ and, consequently, we choose $\lbrace \Ket{\phi_n} \rbrace$ to be identical to the ON eigenbasis $\lbrace \Ket{\psi_n} \rbrace$  of the projected $\hat{H}_\mathrm{rel}$. Hence, we obtain
\begin{equation}
 \hat{\rho}_\mathrm{rel} = \sum_{n=0}^{\mathrm{dim}(\mathcal{F}_\mathrm{rel})-1} w_n \Ket{\psi_n}\Bra{\psi_n}
\end{equation}
which in the Slater determinant basis of natural orbitals $\lbrace \Ket{D_n} \rbrace$ reads as
\begin{equation}
 \rho^\mathrm{D}_{\mathrm{rel}_{kk'}} = \Bra{D_k}\hat{\rho}_\mathrm{rel}\Ket{D_{k'}} = \sum_{n=0}^{\mathrm{dim}(\mathcal{F}_\mathrm{rel})-1} w_n \Lambda_{k'n}^\ast \Lambda_{kn}
\end{equation}
and the $\rho_1$ constraints have the form
\begin{eqnarray}
\label{eq:constraints_B}
 \xi_i\delta_{ij} &\overset{!}{=}& \sum_{n=0}^{\mathrm{dim}(\mathcal{F}_\mathrm{rel})-1} w_n \Bra{\psi_n}c^\dagger_j c_i\Ket{\psi_n}\nonumber\\
 &=& \sum_{n=0}^{\mathrm{dim}(\mathcal{F}_\mathrm{rel})-1} w_n \sum_{k,k'=0}^{\mathrm{dim}(\mathcal{F}_\mathrm{rel})-1} \Lambda_{k'n}^\ast \Lambda_{kn} \Bra{D_{k'}}c^\dagger_j c_i\Ket{D_k}
\quad
(\forall i,j \in\,\mathrm{relevant}).
\end{eqnarray}
For the diagonal coefficients of Eq.~\eqref{eq:diagonal_coeffs} we have
\begin{equation}
 \nu_i^n = \Bra{\psi_n}\hat{n}_i\Ket{\psi_n} = \sum_{k=0}^{\mathrm{dim}(\mathcal{F}_\mathrm{rel})-1} |\Lambda_{kn}|^2 N_i(\Ket{D_k}) \hspace{5mm}.
\end{equation}
The energies $E_n$ associated with each eigenvector $\Ket{\psi_n}$ of $\hat{\rho}_\mathrm{rel}$ are given by the eigenvalues of $\hat{H}_\mathrm{rel}$, that is, $E_n=\epsilon_n$ here. Both $\nu_i^n$ and $E_n$ are required in Eqs.~\eqref{eq:boltzmann} and \eqref{eq:partition} below.

The vector $\mathbf{w}$ of eigenvalues of $\hat{\rho}_{rel}$ with $\mathrm{dim}(\mathcal{F}_\mathrm{rel})=2^{N_\mathrm{rel}}$ components has to satisfy a set of $N_\mathrm{rel}\times N_\mathrm{rel}$ conditions given by Eqs.~\eqref{eq:constraints_B}. If $N_\mathrm{rel}=3$, the number of constraints is larger than the number of unknowns and the problem is overdetermined. If $N_\mathrm{rel}=2~\mathrm{or}~4$, the number of constraints coincides with the number of unknowns. For $N_\mathrm{rel}>4$, the problem is underdetermined, thus requiring further assumptions about the function form of $w_n$, such as a Boltzmann form (see below).

\subsection{Generalized Boltzmann form of $w_n$}

The chosen ansatz of a grand canonical form for the eigenvalues $w_n$ of $\hat{\rho}_\mathrm{rel}$ results from the assumption of maximum entropy, under the average particle number constraints (in natural orbitals) given by Eq.~\eqref{eq:rho1_diagonal_constraint}, the normalization condition and a given temperature. Its form reads as~\cite{jaynes1957}
\begin{equation}
\label{eq:boltzmann}
 w_n = \frac{1}{Z(T,\mu)}\exp\left(-\frac{1}{k_BT}\left(E_n-\sum_{i=0}^{N_\mathrm{rel}-1}\mu_i\nu_i^n\right)\right)
\end{equation}
with the following expression for the partition function
\begin{equation}
\label{eq:partition}
 Z(T,\mu)= \sum_n\exp\left(-\frac{1}{k_BT}\left(E_n-\sum_{i=0}^{N_\mathrm{rel}-1}\mu_i\nu_i^n\right)\right) \hspace{5mm}.
\end{equation}
Here, $\mu_i$ is the effective electrochemical potential of natural orbital $i$ considered as an adjustable free optimization variable, $T$ is the given effective temperature and $k_\mathrm{B}$ is Boltzmann's constant. The partition function ensures normalization, and the condition that $T$ is given is equivalent to fixing the average energy of the system. One has to note that under non-equilibrium conditions (due to an applied voltage $V_\mathrm{DS}$ between the source and drain contacts at the outer ends of the channel), the effective electrochemical potentials $\mu_i$ become independent quantities in general.

\subsection{Numerical determination of $w_n$}

In order to find a solution for the vector $\mathbf{w}$ of $2^{N_\mathrm{rel}}$ eigenvalues of $\hat{\rho}_\mathrm{rel}$, the $N_\mathrm{rel}$ electrochemical potentials $\mu_i$ must be adjusted in such a way that the resulting $\mathbf{w}$ satisfies the $\rho_1$ constraints. Since the problem is underdetermined in general, it is possible that more than a single solution exists. On the other hand, there might exist no exact solution for the constraints at all for the assumed functional form of $w_n$ and eigenbasis of $\hat{\rho}_\mathrm{rel}$. In the latter case, one has to search for an optimum set of $\mu_i$ that minimizes a measure of deviation from the constraint condition.

It can be shown that in case A, the Newton-Raphson method is capable of obtaining solutions that minimize the deviation of the $N_\mathrm{rel}$ differences $\Braket{\hat{n}_i}-\xi_i$ and fulfill condition~\eqref{eq:rho1_diagonal_constraint} very well. Nevertheless, the method cannot be applied to case B, where the number of constraints $N_\mathrm{rel}\times N_\mathrm{rel}$ is larger than the number $N_\mathrm{rel}$ of electrochemical potentials in general.

So for case B, as well as for case A, a single-objective genetic algorithm~\cite{haupt2004} (GA) optimization method is used in the following. Its goal is to minimize an objective function given by the absolute value of the deviation from exact constraint satisfaction with the $\mu_i$ as optimization variables. There are several benefits in using a GA: It operates in parallel with a population of candidate solutions, instead of just a single one. It always yields a solution which improves after every iteration, in contrast to other methods that simply do not give a solution if convergence is not achieved. It can leave local optimum points in the search space behind, even if the objective function is not smooth. We have chosen to implement the GA using the multi-crossover formula described in Ref.~\cite{chang2006}.

\section{Examples}
\label{sec:examples}

As an application of the many-body statistical operator obtained with the method described in the previous sections, we consider the calculation of the electron density and the covariance of the density--density correlation function within the relevant subspace for two different voltage points of the NWFET. For the determination of $\rho_1$, the NEGF/MCSCG approach has been used. For the examples discussed below, we assume the following device parameters: The channel consists of an $\mathrm{InP}$ nanowire of length $L=17.7~\mathrm{nm}$ with $N_\mathrm{sites}=30$ ($N_\mathrm{max}=60$ with spin) and diameter $d_\mathrm{ch}=5~\mathrm{nm}$, the gate oxide consists of $\mathrm{SiO_2}$ with a thickness $d_\mathrm{ox}=10~\mathrm{nm}$, the outer drain and source contacts are assumed to be Schottky contacts with a barrier height of $\Phi_\mathrm{B}=0.7~\mathrm{eV}$, and the temperature is $T=4.2~\mathrm{K}$.

If $\tilde{c}^\dagger_{x\sigma}$ and $\tilde{c}_{x\sigma}$ are the creation and annihilation operators in the localized orbital basis (where the spin is written explicitly as $\sigma$), the electron density at a site $x$ for a given spin $\sigma$ is the expectation value of the operator $\hat{n}_{x\sigma}=\tilde{c}^\dagger_{x\sigma}\tilde{c}_{x\sigma}$. The density--density correlation function is the expectation value of the product of operators $\hat{n}_{x\sigma}\hat{n}_{x'\sigma'}$, and therefore represents a two-particle observable. In turn, the covariance of the density--density correlation function is defined as $\Braket{\hat{n}_{x\sigma}\hat{n}_{x'\sigma'}}-\Braket{\hat{n}_{x\sigma}}\Braket{\hat{n}_{x'\sigma'}}$.

\subsection{Equilibrium case}

In the first example, we consider an equilibrium bias condition, i.e. $V_\mathrm{DS}=0$. The gate voltage $V_\mathrm{GS}=0.35~\mathrm{V}$ is chosen such that the channel is occupied with $N_e=2$ electrons, located within the second Coulomb diamond.

Concerning the structure of the resulting statistical operator in case A, it is dominated by a a single Slater determinant $\Ket{1100}$ with associated weight very close to unity (due to the very low thermal energy). Similarly in case B, $\hat{\rho}_\mathrm{rel}$ is also composed of a single many-body eigenstate with associated unity weight whose only dominant component is the Slater determinant $\Ket{1100}$ with almost unity amplitude $|\Lambda_{00}|^2=1$. In this situation, therefore, we see that case A and B produce (practically) the same statistical operator, corresponding to a pure state to a very good approximation.

\begin{figure}[htb]
 \centering
 \includegraphics[width=0.5\linewidth]{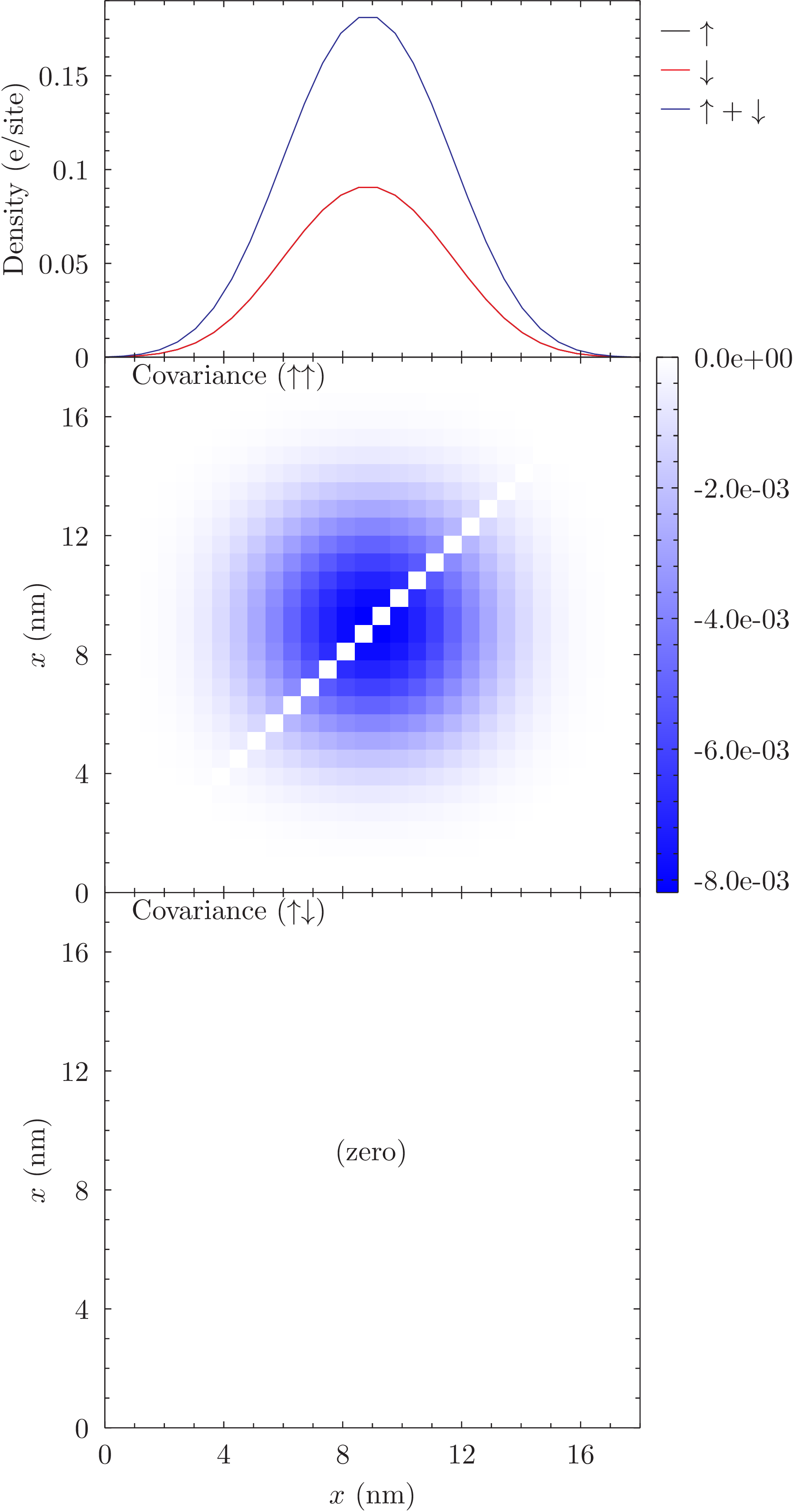}
 \caption{Plots of the electron density and covariance for equilibrium bias condition (identical for case A and B).}
 \label{fig:observables_equilibrium}
\end{figure}

Figure~\ref{fig:observables_equilibrium} gathers the results of the calculated electron density (upper row), the spin up--up covariance (middle row) and spin up--down covariance (lower row) for the chosen equilibrium bias condition. Calculations performed using Slater determinants of natural orbitals (case A) and eigenvectors of $\hat{H}_\mathrm{rel}$ (case B) as eigenbases for $\hat{\rho}_\mathrm{rel}$ yield the same results for the density and covariance. The spin down--down covariance is similar to the spin up--up covariance, since there is no physical reason for spin-symmetry breaking, and is therefore not displayed. For better visualization, we omit to plot the diagonal elements on the graph of the spin up--up covariance, which are given by the expression $\Braket{\hat{n}_{x\sigma}}\left(1-\Braket{\hat{n}_{x\sigma}}\right)$.

One can see that the electron density components for spin up and down are identical. Again, the system preserves spin-symmetry since there is, for example, no applied magnetic field that could break it. The two electrons occupy the same spatial regions with opposite spin, in accordance to Pauli's principle. The spin up--up covariance is negative and has appreciable values around the center of the graph. In contrast, the spin up-down covariance vanishes.

\begin{figure}[htb]
 \centering
 \includegraphics[width=\linewidth]{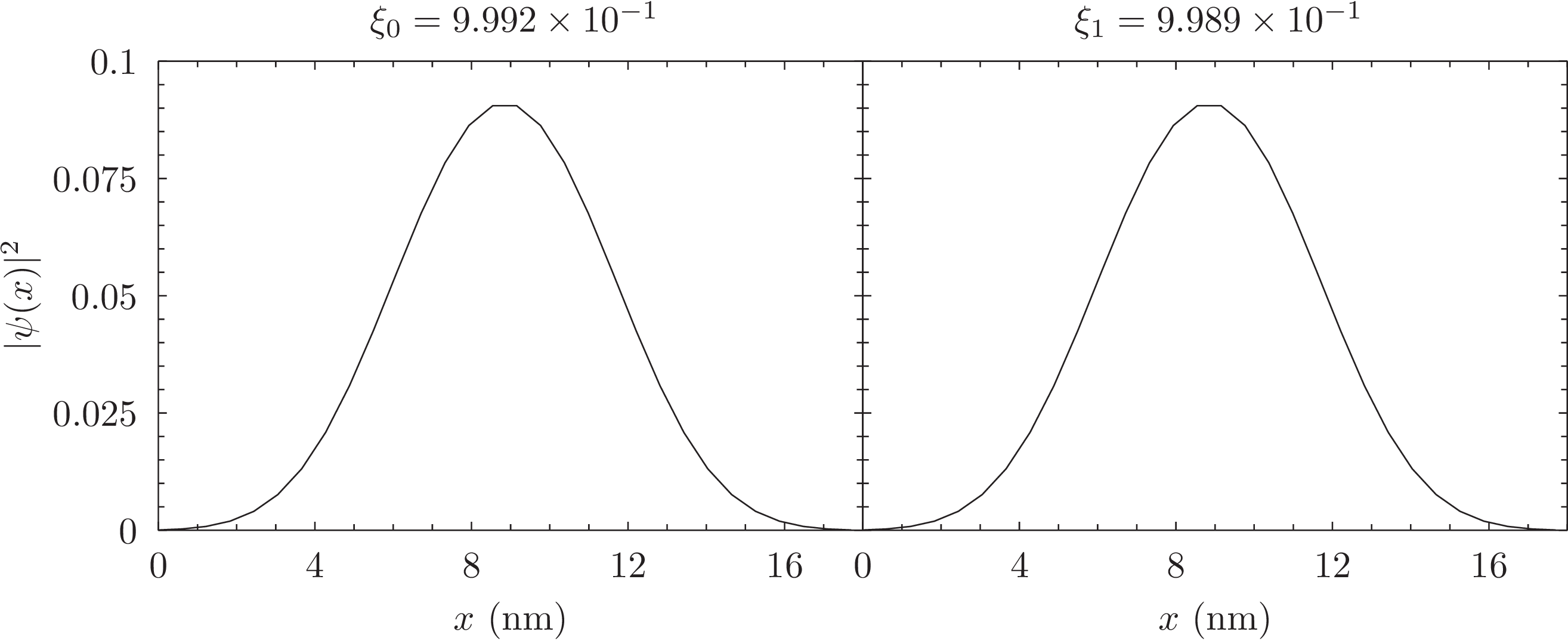}
 \caption{Plots of the modulus squared of the two occupied natural orbital wave functions for equilibrium bias condition. $\xi_i$ is the occupation number of the natural orbital $i$.}
 \label{fig:natorbs_equilibrium}
\end{figure}

The two occupied natural orbitals correspond to the wave functions displayed in Fig.~\ref{fig:natorbs_equilibrium}. We can see that their shape resembles that of the electron density.

\subsection{Non-equilibrium case}

In the second example, we consider a non-equilibrium bias condition, with $V_\mathrm{DS}=0.1~\mathrm{V}$. The gate voltage is the same as in the equilibrium example, $V_\mathrm{GS}=0.35~\mathrm{V}$. In this situation, there are four relevant single-particle states, $N_\mathrm{rel}=4$ and the dimension of the relevant Fock subspace is $\mathrm{dim}(\mathcal{F}_\mathrm{rel})=16$. Here, the statistical operator is not as simple as in the previous example. Table~\ref{tab:nonequilibrium_A} shows the mixture of Slater determinants that compose $\hat{\rho}_\mathrm{rel}$ in case A. The $\Ket{1100}$ determinant is dominant, as in the equilibrium example, but there are contributions from higher excitations and different particle number.

\begin{table}[hbt]
 \centering
  \begin{tabular}{c|c}
   Slater det. & $w_n$ \\
   \hline
   1100 & $5.793\times10^{-1}$ \\
   1010 & $1.861\times10^{-1}$ \\
   0101 & $1.858\times10^{-1}$ \\
   1101 & $2.450\times10^{-2}$ \\
   1110 & $2.432\times10^{-2}$ \\
  \end{tabular}
  \caption{$\hat{\rho}_\mathrm{rel}$ for non-equilibrium case A.}
  \label{tab:nonequilibrium_A}
\end{table}

Table~\ref{tab:projectors_nonequilibrium} shows the probabilities associated with the different Slater determinants, given the resulting statistical operator for case B. (Only the largest contributions are listed.) These probabilities are given by the expectation value of the projectors $\Braket{P_i}=\Braket{\Ket{D_i}\Bra{D_i}}$. In addition, Table~\ref{tab:nonequilibrium_B} shows the individual contributions of Slater determinants to the eigenstates of $\hat{\rho}_\mathrm{rel}$. While there is general accordance with the probabilities of case A, there are some differences. The two Slater determinants $\Ket{0110}$ and $\Ket{1001}$ have appreciable probabilities, in contrast to case A. This is because of the appearance of the many-body state number 5 in the composition of $\hat{\rho}_\mathrm{rel}$ for case B, as seen in Table~\ref{tab:nonequilibrium_B}. A feature that these tables show is the symmetry in the amplitudes and, less pronouncedly, in the weights of those many-body states whose components are mutually complementary. As before, there is no spin-symmetry breaking.

\begin{table}[hbt]
 \centering
  \begin{tabular}{c|c}
   Slater det. & $\braket{P_n}$ \\
   \hline
   1100 & $5.874\times10^{-1}$ \\
   1010 & $1.190\times10^{-1}$ \\
   0101 & $1.177\times10^{-1}$ \\
   0110 & $5.949\times10^{-2}$ \\
   1001 & $5.947\times10^{-2}$ \\
   1101 & $2.518\times10^{-2}$ \\
   1110 & $2.398\times10^{-2}$ \\
   0011 & $7.750\times10^{-3}$ \\
  \end{tabular}
  \caption{Probabilities associated to the Slater determinants.}
  \label{tab:projectors_nonequilibrium}
\end{table}

\begin{table}[hbt]
 \centering
  \begin{tabular}{c|c|c|c}
   MB State & $w_n$ & Slater det. & $|\Lambda_{in}|^2$ \\
   \hline
   \multirow{2}{*}{2} & \multirow{2}{*}{$5.951\times10^{-1}$} & 1100 & $9.870\times10^{-1}$ \\
                      &                                       & 0011 & $1.302\times10^{-2}$ \\
   \hline
   \multirow{3}{*}{4} & \multirow{3}{*}{$1.192\times10^{-1}$} & 1010 & $9.894\times10^{-1}$ \\
                      &                                       & 0110 & $9.072\times10^{-3}$ \\
                      &                                       & 1001 & $1.483\times10^{-3}$ \\
   \hline
   \multirow{4}{*}{5} & \multirow{4}{*}{$1.186\times10^{-1}$} & 0110 & $4.911\times10^{-1}$ \\
                      &                                       & 1001 & $4.911\times10^{-1}$ \\
                      &                                       & 0101 & $8.912\times10^{-3}$ \\
                      &                                       & 1010 & $8.906\times10^{-3}$ \\
   \hline
   \multirow{3}{*}{6} & \multirow{3}{*}{$1.179\times10^{-1}$} & 0101 & $9.894\times10^{-1}$ \\
                      &                                       & 1001 & $9.079\times10^{-3}$ \\
                      &                                       & 0110 & $1.487\times10^{-3}$ \\
   \hline
   \multirow{2}{*}{0} & \multirow{2}{*}{$2.523\times10^{-2}$} & 1101 & $9.722\times10^{-1}$ \\
                      &                                       & 1110 & $2.761\times10^{-2}$ \\
   \hline
   \multirow{2}{*}{1} & \multirow{2}{*}{$2.395\times10^{-2}$} & 1110 & $9.721\times10^{-1}$ \\
                      &                                       & 1101 & $2.761\times10^{-2}$ \\
  \end{tabular}
  \caption{$\hat{\rho}_\mathrm{rel}$ for non-equilibrium case B.}
  \label{tab:nonequilibrium_B}
\end{table}

\begin{figure}[htb]
 \centering
 \includegraphics[width=0.5\linewidth]{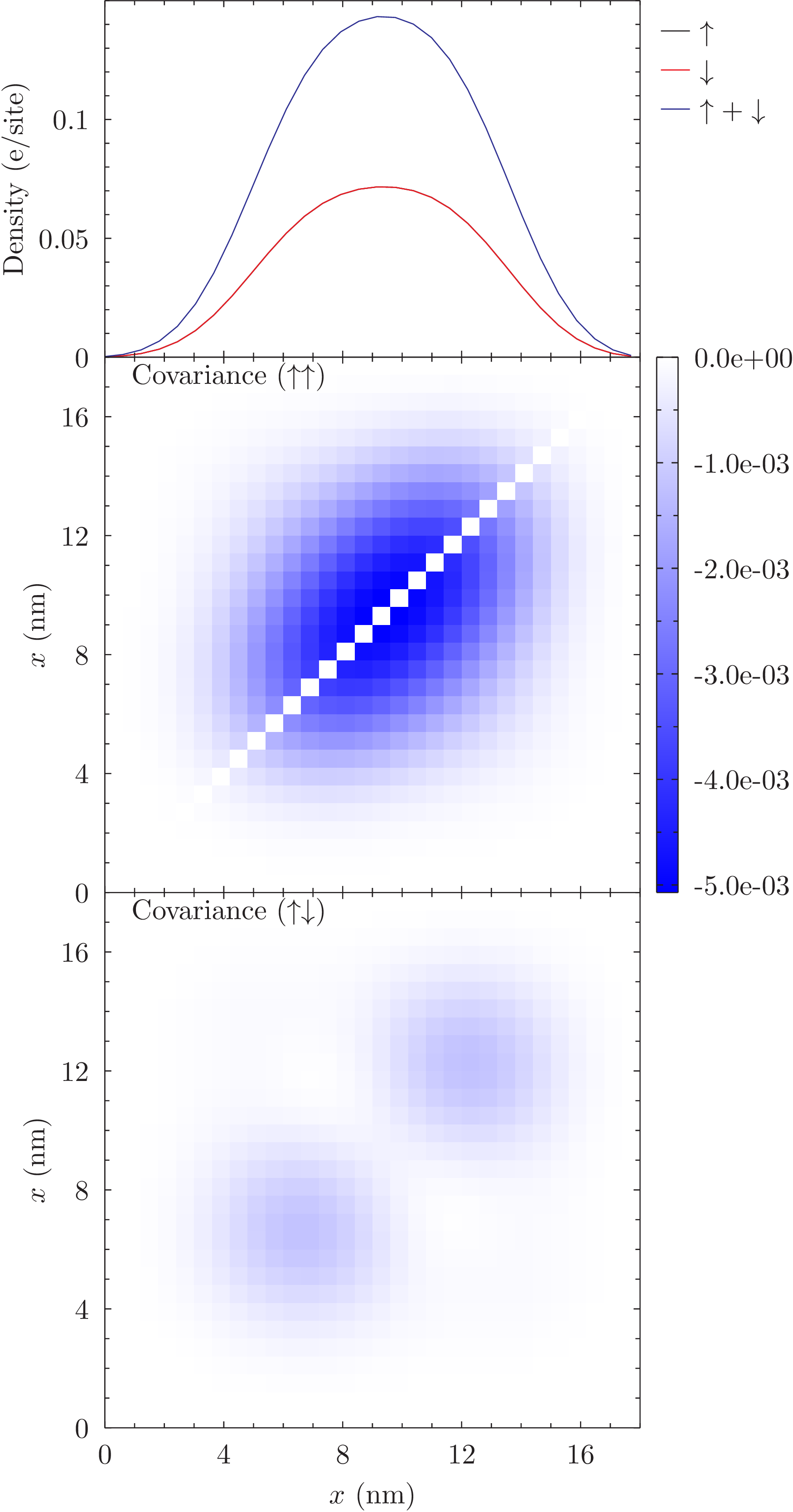}
 \caption{Plots of the electron density and covariance for non-equilibrium bias condition for case B.}
 \label{fig:observables_nonequilibrium}
\end{figure}

\begin{figure}[htb]
 \centering
 \includegraphics[width=\linewidth]{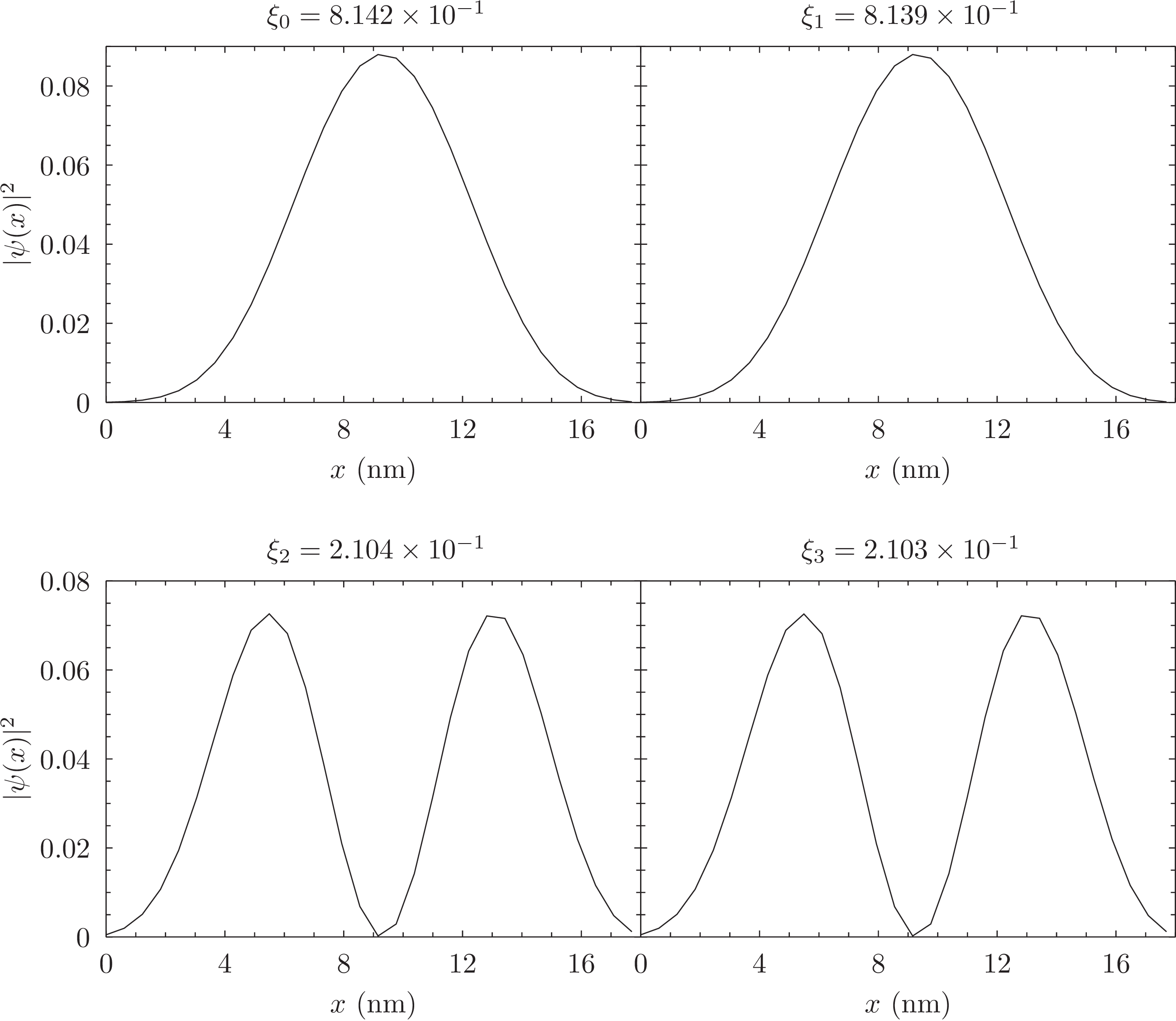}
 \caption{Plots of the modulus squared of the four relevant natural orbital wave functions for non-equilibrium bias condition. $\xi_i$ is the occupation number of the natural orbital $i$.}
 \label{fig:natorbs_nonequilibrium}
\end{figure}

Fig.~\ref{fig:observables_nonequilibrium} shows the resulting electron density and covariance for the non-equilibrium condition for case B, analogous to the previous equilibrium example and the same arguments apply here, except for the non-vanishing spin up--down covariance. The underlying relevant natural orbitals are plotted in Fig.~\ref{fig:natorbs_nonequilibrium}. The plots for case A (Slater determinant eigenbasis) are not show separately since the electron density and natural orbitals are identical to case B (due to identical $\rho_1$), and the deviation in the covariance between case A and case B is smaller that the color scale resolution employed here.

\clearpage

\section{Conclusion}
\label{sec:conclusions}

We have presented a numerical method to obtain the many-body statistical operator $\hat{\rho}_\mathrm{rel}$ for a nanowire-based field-effect transistor as a functional of the single-particle density matrix $\rho_1$, which is obtained by means of the multi-configurational self-consistent Green's function method. In order to make calculations of realistic devices with many basis states feasible, a relevant Fock subspace of Slater determinants of natural orbitals (eigenvectors of $\rho_1$) is employed, based on the classification of these single-particle states into relevant (fluctuating and weakly coupled) and non-relevant. In turn, the relevant subspace is treated in a many-body way, while employing a meanfield approach for the rest. This approach enables a numerical treatment of typical few-electron charging effects for a realistic nanoscale device.

As a constraint, the many-body statistical operator is required to be compatible with the given $\rho_1$. Its eigenvalues $w_n$ are assumed to be of a generalized Boltzmann form, parameterized by electrochemical potentials $\mu_i$ and a given temperature. Two different orthonormal eigenbases of $\hat{\rho}_\mathrm{rel}$ were assumed, corresponding to Slater determinants of natural orbitals or to eigenvectors of the projected many-body Hamiltonian $\hat{H}_\mathrm{rel}$. A genetic algorithm has been employed to determine an optimum set of $\mu_i$ numerically.

In order to demonstrate the applicability of the method, a typical nanowire device example has been presented. With the help of the numerically determined $\hat{\rho}_\mathrm{rel}$, expectation values of observables have been calculated, such as the electron density and the density--density correlation function (and the resulting covariance), for equilibrium and non-equilibrium bias conditions.

\begin{acknowledgments}
 The research leading to these results has received funding from the European Union Seventh Framework Programme under agreement No. 265073 (Nanowiring).
\end{acknowledgments}

\bibliography{jmc_kmi_2014}

\end{document}